\documentclass[epj, nopacs]{svjour}
\usepackage{amsmath,amssymb,graphicx,bm}
\usepackage{color}
\usepackage[bottom]{footmisc}

\newcommand{\beq}{\begin{equation}}
\newcommand{\eeq}{\end{equation}}
\newcommand{\bea}{\begin{eqnarray}}
\newcommand{\eea}{\end{eqnarray}}
\newcommand{\ba}{\begin{align}}
\newcommand{\ea}{\end{align}}
\newcommand{\bfig}{\begin{figure}}
\newcommand{\efig}{\end{figure}}

\newcommand{\D}{\displaystyle}

\newcommand{\thetaprime}{\theta^{\prime}}

\newcommand{\tin}{t_{\rm in}}

\newcommand{\tplus}{t_{+}}

\newcommand{\omnes}{{\cal{O}}}

\newcommand{\ct}{\text{CT}_1}
\newcommand{\ctbar}{\text{CT}_2}

\newcommand{\lprime}{\lambda^{\prime}}
\newcommand{\ldprime}{\lambda^{\prime \prime}}

\begin{document}
\title{Theory of unitarity bounds and low energy form factors}
\author{Gauhar Abbas$^{a}$  \and 
B.\ Ananthanarayan$^{a\,*}$  \and 
I. Caprini$^{b}$  \and I. Sentitemsu Imsong$^{a}$  \and S. Ramanan$^{a}$}
\institute{$^a$ Centre for High Energy Physics,
Indian Institute of 
Science, Bangalore 560 012, India \\
$^b$ National Institute of Physics and Nuclear Engineering,
Bucharest, R-077125, Romania
                                                      \\
{\email{$^*$anant@cts.iisc.ernet.in}}
}        

\date{\today}
\abstract{
We present a general formalism for deriving bounds on the shape parameters of
the weak and electromagnetic form factors using as input correlators
calculated from perturbative QCD, and exploiting analyticity and
unitarity. The values resulting from the symmetries of QCD at low energies or
from lattice calculations at special points inside the analyticity domain can be
included in an exact way. We write down the general solution of the
corresponding Meiman problem  for an arbitrary number of interior constraints
and the integral equations that allow one to include the phase of the form
factor along a part of the unitarity cut. A formalism that includes the phase
and some information on the modulus along a part of the cut is also given.  For
illustration we present constraints on the slope and curvature of the $K_{l3}$
scalar form factor and discuss our findings in some detail. The
techniques are useful for checking the consistency of various inputs and for
controlling the parameterizations of the form factors entering precision
predictions in flavor physics.}

\titlerunning{Unitarity bounds}

\authorrunning{Gauhar Abbas et al.} 

\maketitle
\section{ Introduction\label{sec:intro}}
Form factors are of central importance in strong interaction dynamics,
providing information on the nature of the strong force and
confinement.  Phenomenologically, the weak form factors are of crucial importance for the determination of standard model parameters such as the
elements of the Cabibbo-Kobayashi-Maskawa matrix.

Bounds on the form factors  when a suitable integral of the modulus squared
along the unitarity cut is known from an independent source were considered in
the 1970's  in the context of the hadronic contribution to muon anomaly and the
kaon semileptonic decays
\cite{Okubo,LiPa,Rasz,NeRa,Micu,Micu:1973vp,RaSc,RaScSt,AuMa,Raina,Raina:1978ne}
(for a topical review of the results at that time, see \cite{SiRa}). Through
complex analysis, this condition leads to constraints on the values at interior
points or on the expansion  parameters  around $t=0$,  such as the slope  and
curvature, and belong to a class of problems referred to as the Meiman
problem~\cite{Meiman}. Mathematically, the problem belongs to the standard
analytic interpolation theory for functions in the Hardy class $H^2$
\cite{Meiman,Duren,Raina:1978nd,KrNu}.
The integral condition was provided either from an observable (like muon's $g-2$ in the case of the electromagnetic form factor of the pion), or from the dispersion relation satisfied by a suitable correlator, whose positive spectral function has, by unitarity, a lower bound involving the modulus squared of the relevant form factor.  Therefore, the constraints derived in this framework are often referred to as ``unitarity bounds''.

An important step forward was achieved in \cite{BoMaRa},  where it was noted
that the correlator of interest for the $K_{l3}$ form factors can be
evaluated reliably in the deep Euclidean region by perturbative  QCD. The modern
approach clarified also the issue of the number of subtractions required
in the dispersion relation for the  correlator, which was  not always  treated
correctly in the prior studies. 

The first applications in the modern approach concerned mainly the form factors
relevant for  the $B\to D$ and $B\to D^*$  semileptonic decay, or the so-called
 Isgur-Wise function,  where heavy quark symmetry provided strong
additional constraints at interior points \cite{RaTa,deRafael:1993ib,Lellouch:1995yv,Boyd:1995cf,Boyd:1995sq,CaNe,CaMa,CaLeNe}. More recent
applications revisited the electromagnetic form factor of the pion 
\cite{Caprini2000,arXiv:0801.2023,arXiv:0811.0482,arXiv:0903.4297}, the strangeness changing $K\pi$  form
factors \cite{BC,Hill,arXiv:0905.0951,arXiv:0912.2831}, and the $B\pi$ vector form factor
\cite{BeHi,BoCaLe}. The results confirm that the approach represents a useful
tool in the study of the form factors, complementary  and free of additional
assumptions inherent in standard dispersion relations.

 The  purpose of the present paper is to present in  a systematic way the
technique of unitarity bounds, including its most recent developments and
offering explicit formulas  that can be applied in future studies. There are
several recent developments which increase the interest in these techniques and
justify the present review. First, the correlators used as input are calculated
now with greater precision in perturbative QCD, in many cases up to the order
$\alpha_s^4$. Also, calculations with greater precision in Chiral Perturbation
Theory (ChPT), Heavy Quark Effective Theory (HQET), Soft Collinear Effective
Theory (SCET),  or on the lattice, provide improved values of the form factors
at some specific points. The techniques presented here are the optimal
frame
of including inputs coming from separate sources and testing their consistency.
Moreover, improved information about the phase of the form factor is available
by Watson's theorem \cite{Watson} from the associated elastic scattering, and in
some cases also the modulus is measured independently along a part of the cut. 

We first present in section \ref{sec:notations} our notation and
formalism and in section \ref{sec:meiman} the general Meiman 
interpolation problem, for an arbitrary number of derivatives at the origin and
an  arbitrary number of values at points inside the analyticity  domain. 
The solution is obtained either by Lagrange multipliers, or by the techniques of
analytic interpolation theory \cite{CaDi,Caprini1983}, and is written in two
equivalent ways, as a determinant of a suitable matrix and as a compact convex
quadratic form. 

In section \ref{sec:phase}  we present the
complete treatment of the inclusion of the phase on a part of the unitarity cut,
 along with an arbitrary number of constraints 
of the Meiman type.  No such treatment is found in the 
literature despite the long history of the problem
\cite{Micu,Micu:1973vp,AuMa,Caprini2000,BC}. We derive two equivalent sets of integral
equations, using, as in section \ref{sec:meiman},  either Lagrange multipliers
or the analytic interpolation theory.

In section \ref{sec:modulus} we treat the situation when, in addition
to the phase, some information on the modulus  along a part of the cut is
available from an independent source. This is a mathematically more complicated
problem  \cite{RaScSt,SiRa}.  In this paper we shall present an approach 
proposed  in \cite{Caprini2000}, which 
uses the fact that the knowledge of the phase  allows one to describe exactly
the elastic cut of the form factor by means of the Omn\`{e}s function.  The
problem is thus reduced to a standard Meiman problem  on a larger analyticity
domain.
The method is very powerful and was recently employed to provide stringent 
constraints on the scalar $K\pi$ form factor at low energies
\cite{arXiv:0912.2831}. However, while the method of treating the phase in
section \ref{sec:phase} automatically  takes into account all the constraints of
the original problem, this is not the case with the technique of this section:
it provides necessary constraints for the input, which however may violate the
original unitarity inequality. Therefore, the allowed domain for the input
values is given by the intersection of the domains derived by the formulas of
sections \ref{sec:phase} and \ref{sec:modulus}. We illustrate this fact in section
\ref{sec:example}, where we present constraints  on the slope and curvature of
the $K_{l3}$ scalar form factor. This section extends the results previously
reported in
\cite{arXiv:0912.2831}. Finally, in the last section we summarize our
conclusions and discuss possible applications.

This paper provides a comprehensive treatment of important mathematical
and theoretical tools that are essential for improving precision studies of
form factors that are of great importance to the standard model.

\vskip0.5cm
\section{Notation and Formalism\label{sec:notations}}
 Let $F(t)$ denote a generic form factor, which is real  analytic 
in the complex $t$-plane  cut along the positive real axis from the 
lowest unitarity branch point $\tplus$ to 
$\infty$. The essential condition exploited in the present context 
is  an inequality of the type:
\beq
 \int^{\infty}_{\tplus } dt\ \rho(t) |F(t)|^{2} \leq I,
        \label{eq:I}
\eeq
where $\rho(t)\geq 0$ is a positive semi-definite weight function 
and $I$ is a known quantity. As mentioned in the Introduction, such
inequalities can be obtained starting from  a dispersion relation satisfied by a
suitable correlator,  evaluated  in the deep Euclidean region by perturbative 
QCD, and whose spectral function is  bounded from below by a term involving the
modulus squared of the relevant form factor. 
 An example will be presented in section \ref{sec:example}.

  In the analysis of the semileptonic decays one is interested in the 
parameters of the Taylor expansion at the origin, written as
\beq\label{eq:taylor}
F(t)= F(0) \left[1+ \lprime\, \frac{t}{M^2}  + \ldprime\, \frac{t^2}{2
M^4} 
+ \cdots \right],
\eeq 
where $M$ is a suitable mass and  $\lprime$ and
$\ldprime$ denote the
dimensionless slope and curvature,
 respectively.  Also, from the symmetries of QCD at low energies  and, more
recently, from lattice calculations, one may know $F(t)$ at several special
points inside the analyticity domain.  The standard unitarity bounds exploit
analyticity of the form factor and the inequality (\ref{eq:I}) in order to
correlate in an optimal way these values and the expansion parameters in
(\ref{eq:taylor}).

Additional information on the unitarity cut can be included in the formalism.
According to Watson's theorem \cite{Watson},  below the  
inelastic threshold $\tin$ the phase of $F(t)$  is equal (modulo $\pi$) to the 
phase $\delta(t)$ of the associated elastic scattering process. Thus,
\beq\label{eq:watson}
F(t+i\epsilon)= |F(t)| e^{i\delta(t)}, \quad \quad t_+<t< \tin,
\eeq
where $\delta(t)$ is known. Moreover, in certain cases also some
information on the modulus $|F(t)|$, or a bound on it,  is available on the same
range $t_+<t< \tin$.
In section \ref{sec:meiman} we consider the standard version of the
unitarity bounds, with no information about the phase and the modulus, except
the inequality (\ref{eq:I}). The inclusion of the phase and modulus will
be
discussed in sections \ref{sec:phase} and \ref{sec:modulus}.

For the subsequent treatment, the problem is brought to a canonical form by
making the conformal transformation
\beq\label{eq:z}
z(t)=\frac{\sqrt{t_+}-\sqrt{t_+-t}}{\sqrt{t_+}+\sqrt{t_+-t}}\,,
\eeq
that maps the cut $t$-plane onto the unit disc $|z|<1$ in the $z\equiv z(t)$ 
plane, such that $\tplus$ is mapped onto $z = 1$, the point at 
infinity to $z = -1$ and the origin to $z=0$. After this mapping, the
inequality 
(\ref{eq:I}) is written as
\beq\label{eq:gI}
\frac{1}{2 \pi} \int_{0}^{2\pi} {\rm d}\theta |g(e^{i\theta})|^2 
	\leq I,
\eeq
where the analytic function $g(z)$ is defined as
\beq\label{eq:gz}
g(z) = F(t(z)) w(z).	
\eeq
Here  $t(z)$ is the inverse of (\ref{eq:z}) and  $w(z)$ is an 
{\it outer function}, {\it i.e.}  a function analytic and without zeros in
$|z|<1$, such that its modulus on the boundary is related to $\rho(\theta)$
and the Jacobian of the transformation (\ref{eq:z}). In general, an outer
function is obtained from its modulus on the boundary by the integral
\beq\label{eq:w}
w(z)=\exp\left[\frac{1}{2\pi} \int_{0}^{2\pi} {\rm d}\theta \,
\frac{e^{i\theta}+z}{e^{i\theta}-z}\,\ln |w(e^{i\theta})| \right].
\eeq
In particular cases of physical interest, $w(z)$ has a simple analytic 
form (for an example see section \ref{sec:example}). 

The function $g(z)$ is analytic within 
the unit disc and can be expanded as:
\beq	
g(z)=g_0+g_1 z+ g_2 z^2 + \cdots,
\eeq
and (\ref{eq:I}) implies
\beq\label{eq:gkI}
\sum_{k=0}^\infty g_k^2 \leq I.
\eeq
Using (\ref{eq:gz}), the real numbers $g_k$  are expressed in a straightforward
way  in terms of the coefficients of the Taylor expansion (\ref{eq:taylor}). The
inequality (\ref{eq:gkI}) represents the simplest ``unitarity bound" for the
shape parameters defined in (\ref{eq:taylor}). In what follows we shall improve
it by including additional information on the form factor. 
\vskip0.5cm
\section{ Meiman problem \label{sec:meiman}}
We consider the general case when the first $K$ derivatives of $g(z)$ at $z=0$
and the values at $N$ interior points  are assumed to be known:
\bea\label{eq:cond}
\left[\D \frac{1}{k!} \D \frac{ d^{k}g(z)}{dz^k}\right]_{z=0}&=& g_k, \quad
0\leq k\leq K-1; \nonumber\\
 g(z_n)&=&\xi_n, \quad  1\leq n \leq N, 
\eea
where $g_k$ and $\xi_n$ are given numbers.  They  are related, by (\ref{eq:gz}),
to the derivatives $F^{(j)}(0)$, $j\le k$  of $F(t)$ at $t=0$, and the values
$F(t(z_n))$, respectively. For simplicity
and in view of phenomenological inputs that we will use,
we assume the points $z_n$ to be real, so $\xi_n$ are also real. 

 Meiman problem  \cite{Meiman} requires us to find the optimal
constraints satisfied by the 
numbers defined in (\ref{eq:cond})  if (\ref{eq:gI}) holds.  This mathematical 
problem is also known as a
general  Schur-Carath\'eodory-Pick-Nevanlinna interpolation \cite{Duren,Raina:1978nd,KrNu}.

One can prove \cite{CaDi,Caprini1983} that the most general constraint 
satisfied by the input values appearing in (\ref{eq:cond}) is given by the
inequality:
\beq\label{eq:domain}
\mu_0^2 \leq I,
\eeq
where $\mu_0^2$ is the solution of the minimization problem:
\beq\label{eq:min}
\mu_0^2= \min\limits_{g\in {\cal G}} \, ||g||^2_{L^2}.
\eeq
Here $ ||g||_{L^2}^2$ denotes the $L^2$ norm, {\it i.e.} the quantity appearing 
in the l.h.s. of (\ref{eq:gI}) or (\ref{eq:gkI}), and the minimum is taken over
the class   ${\cal G}$ of analytic functions  
which satisfy  the conditions (\ref{eq:cond}).
In the next subsections we shall solve the minimization problem (\ref{eq:min})
by two different methods.
\vskip0.5cm
\subsection{Lagrange multipliers\label{sec:determinant}}
One may set up a Lagrangian for the minimization problem (\ref{eq:min}) 
with the   constraints (\ref{eq:cond}):
\beq\label{eq:L}
{\cal L} = \frac{1}{2} \sum_{k = 0}^\infty  g_k^2  + \sum_{n=1}^N \alpha_n 
\bar{\xi}_n
\eeq
where $\alpha_n$  are Lagrange multipliers, and $\bar{\xi}_n$ are known numbers
defined as 
\beq
\bar{\xi}_n = \xi_n - \sum_{k=0}^{K-1}g_k z_n^k.
\eeq
Solving the Lagrange equations obtained by varying with respect 
to $g_k$ for all $k \ge  K$, and eliminating the Lagrange  multipliers yields
the solution of the minimization problem (\ref{eq:min}).
For purposes of illustration, when $N = 2$, the Lagrange equations yield
\beq
 \displaystyle g_k=\alpha_1 z_1^k +\alpha_2 z_2^k, \quad \quad k\ge K,
	\label{eq:lagn2}
\eeq
and the solution of the minimization (\ref{eq:min}) takes the form:
\beq
\displaystyle \mu_0^2=\sum_{k=0}^{K-1} g_k^2 +
	\alpha_1 \sum_{k=K}^\infty g_k z_1^k +
	\alpha_2 \sum_{k=K}^\infty g_k z_2^k.
	\label{eq:ineqn2}
\eeq
Then inequality (\ref{eq:domain}) can be expressed in terms of the two Lagrange
multipliers as:
\beq
 \displaystyle 
\alpha_1 \bar{\xi}_1 + \alpha_2 \bar{\xi}_2 \,\leq \,\bar{I},
\label{eq:constn2}
\eeq
where 
\beq\label{eq:tildeI}
\bar{I} = I - \sum_{k = 0}^{K-1} g_k^2
\eeq
 and the constraint conditions themselves are
\bea\label{eq:xi1xi2}
& \displaystyle 
\alpha_1 \frac{z_1^{2 K}}{1-z_1^2} + \alpha_2
\frac{(z_1 z_2)^{K}}{1-z_1 z_2} =\bar{\xi}_1,  \nonumber \\
& \displaystyle 
\alpha_1 \frac{(z_1 z_2)^K}{1-z_1 z_2}+
\alpha_2 \frac{z_2^{2 K}}{1-z_2^2}= \bar{\xi}_2.
\eea
The consistency of eqs. (\ref{eq:constn2}) and (\ref{eq:xi1xi2}) can be written
as:
\beq
\left|
\begin{array}{c c c}
\bar{I} & \bar{\xi}_1 & \bar{\xi}_2 \\
\bar{\xi}_1 & \D \frac{z_1^{2 K}}{1-z_1^2} & \D \frac{(z_1 z_2)^K}{1-z_1z_2} \\
\bar{\xi}_2 & \D \frac{(z_1 z_2)^K}{1-z_1 z_2} & \D \frac{z_2^{2 K}}{1-z_2^2}\\
\end{array}
\right| \geq 0.
\eeq
This can be readily extended to the case of $N$ constraints: 
\beq\label{eq:determinant}
\left|
	\begin{array}{c c c c c c}
	\bar{I} & \bar{\xi}_{1} & \bar{\xi}_{2} & \cdots & \bar{\xi}_{N}\\	
	\bar{\xi}_{1} & \D \frac{z^{2K}_{1}}{1-z^{2}_1} & \D
\frac{(z_1z_2)^K}{1-z_1z_2} & \cdots & \D \frac{(z_1z_N)^K}{1-z_1z_N} \\
	\bar{\xi}_{2} & \D \frac{(z_1 z_2)^{K}}{1-z_1 z_2} & 
\D \frac{(z_2)^{2K}}{1-z_2^2} &  \cdots & \D \frac{(z_2z_N)^K}{1-z_2z_N} \\
	\vdots & \vdots & \vdots & \vdots &  \vdots \\
	\bar{\xi}_N & \D \frac{(z_1 z_N)^K}{1-z_1 z_N} & 
\D \frac{(z_2 z_N)^K}{1-z_2 z_N} & \cdots & \D \frac{z_N^{2K}}{1-z_N^2} \\
	\end{array}\right| \ge 0.
\eeq
Alternatively, the solution
can be obtained by introducing  Lagrange multipliers also for the given
coefficients
$g_k,\, k=0,...,K-1$ in (\ref{eq:L}), as was done in ref. \cite{Raina}.
This leads to the inequality, equivalent to (\ref{eq:determinant}):
\beq
\hspace*{-0.2in}\left|
\begin{array}{c c c  c c c c c c}
I & g_0 & g_1 & \cdots & g_{K-1}&  \xi_1 & \xi_2 & \cdots &\xi_N \\
g_0 & 1 & 0 & 0 & 0   & 1 & 1 & \cdots & 1\\
g_1 & 0 & 1 & 0 & 0 & z_1 & z_2 & \cdots & z_N \\
\vdots & \vdots & \vdots  & \vdots & \vdots & \vdots & \vdots & \vdots
& \\
g_{K-1} & 0 & 0 &  \cdots & 1 & z_1^{K-1} & z_2^{K-1} & \cdots & z_N^{K-1}
\\
\xi_1 & 1 & z_1 &  \cdots & z_1^{K-1}& \frac{1}{1-z_1^2} & \frac{1}{1-z_1 z_2}
& 
\cdots & \frac{1}{1-z_1 z_N}\\
\xi_2 & 1 & z_2  & \cdots & z_2^{K-1}& \frac{1}{1-z_1z_2} & \frac{1}{1- z_2^2}
& 
\cdots & \frac{1}{1-z_2 z_N}\\
\vdots & \vdots  & \vdots & \vdots&\vdots & \vdots & \vdots & \vdots &\\
\xi_N & 1 & z_N &  \cdots & z_N^{K-1}& \frac{1}{1-z_1z_N} & \frac{1}{1- z_2 z_N}
& 
\cdots & \frac{1}{1-z_N^2}\\
\end{array}\right| \ge 0. 
\label{eq:det1}
\eeq
This condition is expressed in a straightforward way in terms of the values of
the form factor $F(t)$ at  $t_i=t(z_i)$ and the derivatives at $t=0$, using eqs.
(\ref{eq:z}) and (\ref{eq:gz}). It can be shown that (\ref{eq:det1}) defines a
convex domain in the space of the input parameters.  
\vskip0.5cm
\subsection{Analytic interpolation theory \label{sec:interp}}

Instead of using Lagrange multipliers, one can implement the constraints
(\ref{eq:cond})  by  expanding the function $g\in {\cal G}$ in the most general
way as \cite{CaDi,Caprini1983}
\beq\label{eq:gtoh}
g(z)= \sum\limits_{k=0}^{K-1} g_k z^k + z^K  \sum\limits_{n=1}^N A_n B_n(z) + 
z^K B_{N+1}(z) h(z).
\eeq
Here the functions $B_n(z)$ are products of Blaschke factors\footnote{A 
Blaschke factor is a function $B(z)$ analytic in $|z|<1$, which satisfies the
conditions $B(z_0)=0$ for a value $|z_0|<1$ and $|B(e^{i\theta})|=1$
\cite{Duren}.} defined recurrently as
\beq\label{eq:Bn}
B_1(z)=1,  B_n(z)=\D \frac{z-z_{n-1}}{1-z z_{n-1}}\,B_{n-1}(z),    2\le n \le
N+1,
\eeq
and the numbers $A_n$ are obtained by solving the system of equations
\beq\label{eq:An}
 \sum\limits_{n=1}^m A_n B_n(z_m)=\D \frac{1}{ z_m^K}\, \left[\xi_m-
\sum\limits_{k=0}^{K-1} 
g_k z_m^k\right],\quad 1\le m\le N,
\eeq
where we took into account the fact that $B_n(z_m)=0$ for $n>m$. We recall that
an expansion equivalent, but slightly different from (\ref{eq:gtoh}), was
proposed in \cite{NeRa}.

The function  $h(z)$ is analytic in $|z|<1$ and is free of constraints.
Expressed
in terms 
of $h$, the minimum norm problem 
(\ref{eq:min}) becomes:
\beq\label{eq:minh}
\mu_0^2= \min\limits_{\{h\}} \, ||h-H||^2_{L^2},
\eeq
where $H$ is a function defined on the boundary of the unit disc
$\zeta=\exp(i\theta$) as
\beq\label{eq:H}
H(\zeta)= -\frac{1}{\zeta^K B_{N+1}(\zeta)} \left[\sum\limits_{k=0}^{K-1} g_k
\zeta^k + 
\zeta^K  
\sum\limits_{n=1}^N A_n B_n(\zeta)\right],
\eeq
and is meromorphic in $|z|<1$.

The solution of (\ref{eq:minh}) is straightforward. We expand:
\beq\label{eq:hH} 
h(\zeta)=\sum\limits_{l=0}^{\infty} h_l \zeta^l, \quad\quad\quad H(\zeta)=
\sum\limits_{l=-\infty}^{\infty} H_l \zeta^l,
\eeq
where $H_l$ are known real numbers defined as
\beq\label{Hl}
H_l=\D \frac{1}{2 \pi i} \int\limits_{|\zeta|=1} \zeta^{-l} H(\zeta) \D 
\frac{{\rm d} \zeta}{\zeta}, \quad\quad\quad -\infty 
\ < l < \infty. 
\eeq
Using the expansions (\ref{eq:hH}), we write (\ref{eq:minh}) as: 
\beq\label{eq:minhl}
\mu_0^2= \min\limits_{\{h_l\}} \, \left[\sum\limits_{l=0}^\infty (h_l-H_l)^2 + 
\sum\limits_{l=-\infty}^{-1} H_l^2 \right],
\eeq 
where the minimization is  taken upon the  numbers $h_l$. The minimum 
is reached for 
\beq
h_l=H_l, \quad\quad l\ge 0,
\eeq
which leads to the minimum
\beq\label{eq:minHl}
\mu_0^2=  \sum\limits_{l=-\infty}^{-1} H_l^2.
\eeq
The coefficients $H_l$ for $l\le -1$ are calculated by inserting into (\ref{Hl})
the
expression 
of $H$ from (\ref{eq:H}) and applying the residues 
theorem.  The poles are produced by the factors $B_{N+1}$ and
$\zeta^{K+l-k}$, for $K+l-k\ge 0$. The contribution of $B_{N+1}$ 
to $H_l$ is written as,
\beq\label{eq:Hl1}
 -\sum\limits_{n=1}^N \left[\frac{z-z_n}{B_{N+1}(z)}\right]_{z=z_n}
\hspace{-0.6cm} z_n^{-l-1}\left[ 
\sum\limits_{k=1}^{K-1} g_k z_n^{k-K} + \sum\limits_{m=1}^n A_m B_m(z_n)\right],
\eeq
where in the parenthesis we recognize from (\ref{eq:An}) the numbers
$\xi_n/z_n^K$. 
The factor  $\zeta^{K+l-k}$ contributes to $H_l$ 
as
\beq\label{eq:Hl2}
- \sum\limits_{k=0}^{K-1}\theta(K+l-k)\, \frac{g_k}{(K+l-k)!} 
\frac{{\rm d}^{K+l-k}}{{\rm d}z^{K+l-k}}\,\left[\D 
\frac{1}{B_{N+1}(z)}\right]_{z=0}.
\eeq
Collecting the terms (\ref{eq:Hl1}) and  (\ref{eq:Hl2}) we obtain, for $ l\le
-1$,
\beq\label{eq:Hl12}
H_l= -\sum\limits_{n=1}^N \frac{Y_n\xi_n }{z_n^{l+K+1}}-
\sum\limits_{k=0}^{K-1} 
\theta(K+l-k)  \ g_k \beta_{kl}, 
\eeq
where we denoted 
\bea\label{eq:Ynbetakl}
 Y_n & =&\left[ \frac{z-z_n}{B_{N+1}(z)}\right]_{z=z_n}, \nonumber\\\beta_{kl}&
= &\D 
\frac{1}{(K+l-k)!} \D \frac{{ d}^{K+l-k}}{{ d}z^{K+l-k}}\,
\left[\frac{1}{B_{N+1}(z)}\right]_{z=0}.
\eea
Using (\ref{eq:Hl12})  it is easy to calculate the sum 
required in (\ref{eq:minHl}). Due to the $\theta$ function, only 
the values $l\ge k-K$  in the second term of $H_l$ give non-vanishing 
contributions. The result is written in a compact form as:
\beq\label{eq:mu02}
\mu_0^2=\hspace{-0.2cm}\sum\limits_{m,n=1}^N {\cal A}_{mn} \xi_n \xi_m +
\sum\limits_{j,k=0}^{K-1} 
{\cal B}_{jk} g_j g_k +  2 
\sum\limits_{n=1}^{N}\sum\limits_{k=0}^{K-1} {\cal C}_{kn} g_k \xi_n,
\eeq
where we defined
\bea
{\cal A}_{mn}&=&  \frac{Y_n Y_m}{ z_n^K z_m^K}\,\D \frac{1}{1-z_n z_m}, \quad
{\cal B}_{jk}= \sum\limits_{l=L}^{-1}\beta_{jl}\beta_{kl},\nonumber\\
{\cal C}_{kn}&=& \frac{Y_n}{z_n^K}\, \sum\limits_{l=k-K}^{-1} \frac{\beta_{kl}}{
z_n^{l+1}},
\eea
and  $L=\max (k-K,\, j-K)$. Inserting  (\ref{eq:mu02}) into the inequality 
(\ref{eq:domain}) gives the allowed domain of the input values appearing in
(\ref{eq:cond}):
\beq\label{eq:domain1}
\sum\limits_{m,n=1}^N \hspace{-0.2cm} {\cal A}_{mn} \xi_n \xi_m +
\sum\limits_{j,k=0}^{K-1}
{\cal B}_{jk} 
g_j g_k +  2 
\sum\limits_{n=1}^{N}\sum\limits_{k=0}^{K-1} {\cal C}_{kn} g_k \xi_n \leq I.
\eeq
It can be checked that the domains given by (\ref{eq:determinant}) and 
(\ref{eq:domain1}) are equivalent.

\vskip0.5cm
\section{Inclusion of the phase   \label{sec:phase}}
In this section we shall impose the condition that the phase of the form 
factor is known (modulo $\pi$) along the elastic part of the cut from the phase
of the associated elastic amplitude by Watson's theorem \cite{Watson}.   We start
by defining the
Omn\`{e}s function
\beq	\label{eq:omnes}
 \omnes(t) = \exp \left(\D\frac {t} {\pi} \int^{\infty}_{\tplus} dt 
\D\frac{\delta (t^\prime)} {t^\prime (t^\prime -t)}\right),
\eeq
where $\delta(t)$  is  known for 
$t\le \tin$, and is an arbitrary function, sufficiently  smooth ({\em i.e.}
Lipschitz continuous) for $t>\tin$. From  (\ref{eq:watson}) and (\ref{eq:omnes})
it follows that
\beq\label{eq:ratio}
{\rm Im} \left[\frac{ F(t + i\epsilon)}{ \omnes(t +i\epsilon )}\right]=0, 
\quad\quad  t_+\le t \le \tin.
\eeq
 Expressed in terms of the function $g(z)$ this condition becomes
\beq\label{eq:img}
{\rm Im} \left[\frac{g (e^{i\theta})} {W(\theta)}\right] =0, \quad \quad 
\theta \in (-\theta_{\rm in},\theta_{\rm in}).
\eeq
Here $\theta_{\rm in}$ is defined by $z(\tin)=\exp(i\theta_{\rm in})$ and  
the function $W(\theta)$ is defined as:
\beq\label{eq:W}
W(\theta)= w(e^{i\theta} ) O(e^{i\theta}),
\eeq 
where  $w(z)$  is the outer function and 
\beq\label{eq:O}
O(z)=\omnes(t(z)). 
\eeq
 As shown in \cite{Micu,Caprini2000,BC}, the constraint (\ref{eq:img}) can be 
imposed by means of a generalized Lagrange multiplier. The
constraints at interior points can be treated either with Lagrange
multipliers as in subsection \ref{sec:determinant}, or by their explicit
implementation as in subsection \ref{sec:interp}. Below we shall briefly present
these two approaches.

\vskip0.5cm
\subsection{Lagrange multipliers \label{sec:lagrange}}

The Lagrangian of the minimization problem (\ref{eq:min}) with the constraints
(\ref{eq:cond}) and (\ref{eq:img}) reads
\bea
{\cal L} &=& \frac{1}{2} \sum_{k = 0}^{\infty} g_k^2 + \sum_{n=1}^N
\alpha_n (\xi_n - \sum_{k = 0}^{\infty} g_k z^k)\\ 
&+& \frac{1}{\pi} \sum_{k = 0}^{\infty} g_k \lim_{\rm r \rightarrow 1} 
\int\limits_{-\theta_{\rm in}}^{\theta_{\rm in}} \lambda(\theta') 
|W(\theta')| {\rm \text{Im}} [[W(\theta')]^{-1} r^k e^{i k \theta'}] 
d\theta'.\nonumber 
\eea
The Lagrange multiplier $\lambda(\theta)$ is an odd function, 
$\lambda(-\theta) = -\lambda(\theta)$ and, as in \cite{Caprini2000,BC}, 
the factor  $|W(\theta)|$ was introduced for convenience.

We  minimize ${\cal L}$ by brute force method with respect to the free 
parameters $g_k$ with $k\geq K$. The Lagrange multipliers $\lambda(\theta)$ and
$\alpha_n$  are  found in the standard way by imposing the
constraints (\ref{eq:cond}) and (\ref{eq:img}).  This leads to a system of
coupled equations, which can be solved numerically.

The calculations are straightforward (see for instance \cite{Caprini2000}). In order 
to write the equations in a simple form, it is convenient to define the phase
$\Phi(\theta)$ of the function $W(\theta)$ by
\beq
W(\theta) = |W(\theta)| e^{i \Phi(\theta)}.
\label{wtheta_eqn1}
\eeq
From (\ref{eq:W}) we have
\beq
\Phi(\theta) = \phi(\theta) + \delta (t(e^{i\theta})),
\label{phi_eqn1}
\eeq 
where $\phi(\theta)$ is the phase of the outer function $w(e^{i\theta})$ 
and $\delta(t)$ is the  elastic scattering phase shift. We introduce also the
functions $\beta_n$ for $n=1,...N$, by
\beq\label{eq:betan}
\beta_n(\theta)=z_n^{K} \,\frac{\sin[K \theta-\Phi(\theta)]- 
\sin[(K-1)\theta-\Phi(\theta)]}{1+z_n^2-2 z_n\cos\theta}.
\eeq
Then the  equations for the Lagrange multipliers $\lambda(\theta)$ and $\alpha_n$ 
take the form:
\bea\label{eq:eq1}
&& \sum_{k = 0}^{K-1} g_k \sin[k \theta - \Phi(\theta)] =\lambda(\theta) 
-\sum_{n=1}^N \alpha_n \beta_n(\theta)   \\                     
&&\hspace{-0.1cm} - \frac{1}{2\pi} \int\limits_{-\theta_{\rm in}}^{\theta_{\rm
in}} 
{\rm d} \thetaprime \lambda(\thetaprime) ) {\cal K}_{\Phi}(\theta, \thetaprime),
\quad \quad \theta \in (-\theta_{\rm in},\theta_{\rm in}),  \nonumber 
\eea
\beq\label{eq:eq2}
- \frac{1}{\pi} \int\limits_{-\theta_{\rm in}}^{\theta_{\rm in}} 
\lambda(\theta)\beta_n(\theta) {\rm d}\theta 
+ \sum_{n'=1}^N \alpha_{n'}  \frac{(z_n z_{n'})^{K}}{1-z_nz_{n'}} = \bar{\xi}_n, 
\eeq
where  $n=1,\ldots N$.  

The integral kernel in (\ref{eq:eq1}), defined as
 \beq\label{eq:calK}
{\cal K}_{\Phi}(\theta, \thetaprime) \equiv \frac{\sin[(K-1/2) (\theta -
\thetaprime) - \Phi(\theta) +
\Phi(\theta^\prime)]}{\sin[(\theta-\theta^\prime)/2]}, 
\eeq
  is of Fredholm type if the phase $\Phi(\theta)$ is Lipschitz continuous \cite{BC}. 
Then the  above system can be solved numerically in a straightforward manner. 
Finally, the inequality (\ref{eq:domain}) takes the form:
\beq
\label{eq:domain2}
\frac{1}{\pi} \sum_{k = 0}^{K-1} g_k \int\limits_{-\theta_{\rm
in}}^{\theta_{\rm
in}} 
{\rm d} \theta \lambda(\theta) 
\sin\left[k \theta - \Phi(\theta) \right] +\sum_{n=1}^N \alpha_n \bar{\xi}_n \leq \bar{I},
\eeq
with $\bar I$ defined in (\ref{eq:tildeI}). Using the relation (\ref{eq:gz}), 
the above inequality defines an allowed domain for the values of the form factor 
and its derivatives at the origin. 
Note that removing the phase constraint gives back the results of
subsection \ref{sec:determinant}. The results when phase alone is included
($N = 0$) as in \cite{Caprini2000} (and references therein) and the case $N = 1$ discussed
in \cite{BC,arXiv:0811.0482} are also readily reproduced. It must be emphasized
that the theory for arbitrary number of constraints is being presented here
for the first time.

It is easy to see that, if $\tin$ is increased, the allowed domain defined by the 
inequality  (\ref{eq:domain2}) becomes smaller. The reason is that by increasing 
$\tin$  the class of functions entering the minimization (\ref{eq:min}) becomes 
gradually smaller, leading to a larger value for minimum $\mu_0^2$ entering the 
definition (\ref{eq:domain}) of the allowed domain.

\vskip0.5cm
\subsection{Analytic interpolation theory \label{sec:int}}
 Alternatively, we shall  implement the constraints in 
the interior points by expressing the function $g(z)$  as in eq. (\ref{eq:gtoh}) of subsection 
\ref{sec:interp}, in terms of a function $h(z)$ free of constraints.  The Lagrangian  
can be expressed in terms of the
coefficients $h_l$ defined in (\ref{eq:hH}) as:
\beq\label{eq:Lhl}
{\cal L}= \sum\limits_{l=0}^\infty (h_l-H_l)^2 + \sum\limits_{l=-\infty}^{-1} 
H_l^2 - 2\sum\limits_{l=0}^\infty h_l\, c_l +\cdots
\eeq
where
\beq\label{eq:cl}
c_l= \frac{1}{\pi} \int\limits_{-\theta_{\rm in}}^{\theta_{\rm in}} {\rm
d}\theta'  
\lambda(\theta') \,|W(\theta')|\, {\rm Im} 
\left[ \frac{e^{i \theta' (K+l)} B_{N+1}(e^{i\theta'})}{W(\theta')}\right].
\eeq
The minimization of the Lagrangian given in (\ref{eq:Lhl})  with respect to $h_l$
has 
the simple solution
\beq\label{eq:hlopt}
h_l=H_l + c_l, \quad\quad\quad l\ge 0
\eeq
 leading to the minimum
\beq\label{eq:mu0mod}
\mu_0^2= \sum\limits_{l=0}^\infty c_l^2 +  \sum\limits_{l=-\infty}^{-1} H_l^2.
\eeq
The second sum in the r.h.s. was already evaluated in subsection
\ref{sec:interp}. 
The first term involves the function $\lambda(\theta)$, which we determine   by
imposing the condition (\ref{eq:img}),  written in terms of
  the function $h(z)$  expanded as in  (\ref{eq:hH}). Using $h_l$  given
by (\ref{eq:hlopt}) as the sum $H_l+c_l$, we obtain
by a straightforward calculation  the integral equation for a function  
$\lambda(\theta)$ in the interval $\theta \in (-\theta_{\rm in},\theta_{\rm in})$:
\beq\label{eq:inteq}
\lambda(\theta) - \frac{1}{2 \pi}  \int\limits_{-\theta_{\rm in}}^{\theta_{\rm
in}} 
{\rm d}\theta'  \lambda(\theta') 
 {\cal K}_\Psi(\theta, \thetaprime) = V(\theta),
\eeq
where the kernel ${\cal K}_\Psi$ is defined as in (\ref{eq:calK}) in terms of the known function
\beq
\Psi(\theta)=\arg[B_{N+1}(\exp(i\theta))]-\arg[W(\theta)],
\eeq
and  $V$ is a known function defined as:\bea\label{eq:V}
 V(\theta)&=& \sum\limits_{n=1}^N \frac{Y_n\xi_n}{ z_n^K}\,
\D \frac{z_n\sin[\Psi(\theta)]- \sin[\Psi(\theta)-\theta]}{1+z_n^2-2 z_n \cos\theta}  \nonumber \\&+& \sum\limits_{k=0}^{K-1} \D \frac{g_k}{(K-k-1)!} \, \left[\D \frac{{\rm d}^{K-k-1}U(z)} {{\rm d}z^{K-k-1}}\right]_{z=0},
\eea
with \beq
U(z)=\frac{z\sin[K\theta+\Psi(\theta)] -\sin[(K-1)\theta+\Psi(\theta)]} {B_{N+1}(z) (1+z^2-2 z\cos\theta)}.
\eeq
Using the expression (\ref{eq:cl}) of $c_l$ and the integral equation
(\ref{eq:inteq}), 
it is straightforward to 
evaluate the sum in the first term of (\ref{eq:mu0mod}):
\beq\label{eq:sumcl2}
\sum\limits_{l=0}^\infty c_l^2 =\D \frac{1}{\pi} \int\limits_{-\theta_{\rm in}}^{\theta_{\rm in}} {\rm d}\theta  \lambda(\theta) V(\theta).
\eeq
 Collecting all the terms in (\ref{eq:mu0mod}), (\ref{eq:domain}) can be written as:
\bea\label{eq:domain3}
 \frac{1}{\pi} \int\limits_{-\theta_{\rm in}}^{\theta_{\rm in}} {\rm d}\theta  \lambda(\theta) V(\theta)&+&\sum\limits_{m,n=1}^N 
{\cal A}_{mn} \xi_n \xi_m + \sum\limits_{j,k=0}^{K-1} {\cal B}_{jk} g_j g_k\nonumber
\\ &+& 2 \sum\limits_{n=1}^{N}\sum\limits_{k=0}^{K-1} 
{\cal C}_{kn} g_k \xi_n\, \le I.
\eea
This inequality gives the allowed domain for the input values appearing in the
conditions 
(\ref{eq:cond}) when the phase is known in the elastic region.  We note that the
arbitrary function $\delta(t)$ for $t >\tin$ entering the Omn\`es function
(\ref{eq:omnes}) does not appear in the result. The first term in
(\ref{eq:domain3})  represents the improvement brought by the information on the
phase, as can be seen  by comparing with (\ref{eq:domain1}).  It can be checked
numerically that the allowed domains described by  (\ref{eq:domain2}) and
(\ref{eq:domain3}) are equivalent.
\vskip0.5cm

\section{Inclusion of phase and modulus  \label{sec:modulus}}
In some cases, information on the modulus of the form factor along an interval of 
the unitarity cut is available from an independent source. As we mentioned in the 
Introduction, a rigorous implementation of this information is difficult. 
In this section, we shall present an approach proposed in \cite{Caprini2000}, 
which leads to an independent constraint that should be satisfied by the inputs.

We start with  the remark that   the knowledge of the phase was
implemented in the previous section by the relation 
(\ref{eq:ratio}), which says that the function $f(t)$ defined through
\beq
F(t) = f(t) \omnes(t),
\eeq
is real in the elastic region. In fact, since the Omn\`{e}s function 
$\omnes(t)$ 
fully accounts for  the elastic cut of the form factor,  the function $f(t)$ has
a larger analyticity domain, namely the complex $t$-plane cut only for $t>\tin$.
 Implementing this fact leads to a modified version of the unitarity bounds,
proposed  in \cite{Caprini2000}. The method requires also the
modulus of the form factor in the elastic region. Indeed, (\ref{eq:I}) implies
that $f$ satisfies the  condition 
\beq\label{eq:fI}
\int_{\tin}^\infty {\rm d}t \rho(t) |\omnes(t)|^2 |f(t)|^2 \leq I^\prime
\eeq
where
\beq\label{eq:I1}
I^\prime= I - \int^{\tin}_{\tplus} {\rm d}t \rho(t) |F(t)|^2.
\eeq
If the modulus $|F(t)|$ is known for $\tplus \le t \le \tin$, the quantity
$I^\prime$ 
is known. Then (\ref{eq:fI}) leads, through the techniques presented in section
\ref{sec:meiman}, to constraints on the values of $f$  inside the analyticity
domain.

The problem is brought into the canonical form by the transformation
\beq\label{eq:ztin}
\tilde z(t) = \frac{\sqrt{\tin}-\sqrt {\tin -t} } {\sqrt {\tin}+\sqrt {\tin -t}}\,,
\eeq
which maps the complex $t$-plane cut for $t>\tin$ on
to the unit disc in the $z$-plane defined by $z=\tilde z(t)$. Then
(\ref{eq:fI}) 
can be written as
\beq\label{eq:gI1}
\frac{1}{2 \pi} \int^{2\pi}_{0} {\rm d} \theta |g(\exp(i \theta))|^2 \leq I^\prime,
\eeq
where  the function $g$ is now
\beq\label{eq:gtilde}
 g(z) = \tilde{w}(z)\, \omega(z) \,F(\bar t(z)) \,[O(z)]^{-1}.
\eeq 
Here  $\tilde{w}(z)$ is the outer function related to the weight $\rho(t)$ and 
the Jacobian of the new mapping (\ref{eq:ztin}) and $O(z)$ is defined as
\beq\label{eq:Otilde}
O(z) = \omnes(\tilde t(z)),
\eeq
where  $\tilde t(z)$ is the inverse of $z=\tilde z(t)$ with $\tilde z(t)$ defined in (\ref{eq:ztin}), and
\beq\label{eq:omegatin}
 \omega(z) =  \exp \left(\D\frac {\sqrt {\tin - \tilde t(z)}} {\pi} \int^{\infty}_{\tin} {\rm d}t^\prime \D\frac {\ln |\omnes(t^\prime)|}
 {\sqrt {t^\prime - \tin} (t^\prime -\tilde t(z))} \right).
\eeq 

The inequality (\ref{eq:gI1}) has exactly the same form as (\ref{eq:gI}).
Therefore, 
by using the techniques of section \ref{sec:meiman} we derive constraints on the
function $g$ at  interior points, which can be written in the equivalent forms as in 
(\ref{eq:determinant}) or (\ref{eq:domain1}). Using (\ref{eq:gtilde}), these
constaints are expressed in terms of the physically interesting values of the
form factor $F(t)$. 

In fact,
the Omn\`es function $\omnes(t)$ defined in (\ref{eq:omnes}) is not unique, 
as it
involves the arbitrary function $\delta(t)$ for $t>\tin$.  We have seen
that the results of section \ref{sec:phase} are not affected by this
arbitrariness. 
This is true also for the results of this section: the reason is that a change
of the function  $\delta(t)$ for $t>\tin$ is equivalent with a multiplication of
$g(z)$ by a function analytic and without zeros  in $|z|<1$ ({\it i.e.} an outer
function). According to the general theory of analytic functions of Hardy class \cite{Duren,KrNu}, the
multiplication by an outer function does not change the class of functions used
in minimization problems. In our case,  the arbitrary function
$\delta(t)$ for $t>\tin$ enters in both the functions $O(z)$ and  $\omega(z)$
appearing in (\ref{eq:gtilde}), and their ambiguities compensate each other
exactly. 
The independence of the results on the choice  of the phase  for $t>\tin$ is
confirmed  numerically, for functions $\delta(t)$ that are Lipschitz
continuous. It is important to emphasize that the method relies on the
Omn\`{e}s function making an appearance first through a related outer function
and then through its inverse, while the 
function $f(t)$ is merely introduced at an
intermediate stage and is subsequently eliminated.

The constraints provided by the technique of this section are expected to be
quite strong since they result from a minimization on a restricted class of
analytic functions, where the second Riemann sheet of the form factor is
described explicitly  by the Omn\`es function. On the other hand, it is easy to
see that the fulfillment of the condition (\ref{eq:fI}) does not automatically
imply that the original condition (\ref{eq:I}) is satisfied.  More
exactly, the technique described here   does not impose the knowledge of the
modulus in addition and simultaneously with the  bound (\ref{eq:I}) and the
knowledge of the phase, but exploits only a consequence of the original
conditions of the problem. Therefore, one must calculate separately the allowed
domains of the parameters of interest  given by the techniques of sections
\ref{sec:phase} and \ref{sec:modulus}, and take as the final results the
intersection
of these domains.

\vskip0.5cm
\section{Example: scalar $K\pi$ form factor \label{sec:example}}
We consider as an example the scalar  $K\pi$ form factor $f_0(t)$, presenting
constraints on the slope  $\lprime_0$ and curvature $\ldprime_0$, appearing in
the expansion 
\beq\label{eq:taylor1}
f_0(t)= f_0(0) \left[1+ \lprime_0\, \frac{t}{M^2_\pi}  + \ldprime_0\,
\frac{t^2}{2
M^4_\pi} 
+ \cdots \right],
\eeq 
often used in the physical range of the semileptonic decay $K\to\pi l\nu$. We work in the isospin limits, 
adopting the convention that $M_K$ and $M_\pi$ are the masses of the charged mesons.

 The scalar $K\pi$  form factor has been calculated at low energies in ChPT and
on the lattice (for recent reviews see \cite{arXiv:0801.1817,LL2009,HL2009}). 
At $t=0$, the current value  $f_0(0)=0.962\pm 0.004 $ \cite{LL2009} shows
that the corrections to the  Ademollo-Gatto theorem are quite small. 
Other low energy theorems frequently used are \cite{CallanTreiman,Oehme}
\beq\label{eq:CT}
f_{0}(\Delta_{K \pi}) =\frac{F_K}{F_\pi} + \Delta_{CT}, \quad  f_{0}(\bar\Delta_{K\pi}) =\frac{F_\pi}{F_K} + \bar\Delta_{CT}, \quad 
\eeq
where $\Delta_{K\pi} = M^{2}_{K}-M^{2}_{\pi}$ and  $\bar\Delta_ {K\pi} = -\Delta_{K\pi}$ are the  first and second Callan-Treiman points, denoted below as $\ct$ and $\ctbar$, respectively. The lowest order values are known from $F_K/F_\pi =1.193 \pm 0.006$  \cite{LL2009}, and the corrections calculated to one loop are $\Delta_{CT}=-3.1\times 10^{-3}$  and  $\bar \Delta_{CT}=0.03$  \cite{GaLe19852}. In the isospin limit, the higher corrections are negligible at the first point, but are expected  to be quite large  at the second one.

 An inequality of the type (\ref{eq:I}) is obtained for  $f_0(t)$ starting with
a dispersion relation
satisfied by a suitable correlator of the strangeness-changing current
\cite{BC,Hill}:
\beq\label{eq:Pi0} 
\chi_{_0}(Q^2)\equiv \frac{\partial}{ \partial q^2} \left[ q^2\Pi_0 \right] 
= \frac{1}{\pi}\int_{t_+}^\infty\!dt\, \frac{t {\rm Im}\Pi_0(t)}{ (t+Q^2)^2},
\eeq 
where unitarity implies the inequality: 
\beq\label{eq:unit}
{\rm Im} \Pi_0(t) \ge \frac{3}{2} \frac{t_+ t_-}{ 16\pi} 
\frac{[(t-t_+)(t-t_-)]^{1/2}}{ t^3} |f_0(t)|^2  \,,
\eeq
with $t_\pm=(M_K \pm M_\pi)^2$. 
 The quantity $\chi_{_0}(Q^2)$  was calculated up to  the order $\alpha_s^4$ in perturbative QCD  \cite{BCK2006}. From (\ref{eq:Pi0}) and (\ref{eq:unit}) it follows that in this case the quantity $I$ appearing in (\ref{eq:gI}) is
\beq\label{eq:I2}
I=\chi_0(Q^2).
\eeq
For illustration we give  also the outer function $w(z)$ entering (\ref{eq:gz}):
\bea	\label{eq:w1}
 w(z)&=& \frac{\sqrt{3}}{32 \sqrt{\pi}} \,\frac{M_K-M_\pi}{M_K+M_\pi}\,(1-z)\,(1+z)^{3/2} \nonumber \\
&& \times\,\frac{ (1+z(-Q^2))^2} {(1-z\,z(-Q^2))^2}\frac{(1-z\, z(t_-))^{1/2}}{(1+ z(t_-))^{1/2}},
\eea
Further,   below the inelastic threshold  $\tin$ the phase is known from the $I=1/2$ $S$-wave of
$K\pi$ elastic scattering \cite{BuDeMo}, while the modulus $|f_0(t)|$ was measured recently from the decay
$\tau\to K \pi\nu_\tau$  \cite{Belle}. For details of the input quantities see \cite{arXiv:0912.2831}.
When the modulus is also included, as in section (\ref{sec:modulus}), the
outer functions in (\ref{eq:gtilde}) can be written as, 
\bea	\label{eq:wtil1}
 \tilde{w}(z)&=& \frac{\sqrt{3}(M_K^2-M_\pi^2)}{16 \sqrt{2\pi} \tin}
 \frac{\sqrt{1-z}\,(1+z)^{3/2} (1+\tilde{z}(-Q^2))^2} {(1-z\,\tilde{z}(-Q^2))^2}
\nonumber \\
&& \times\,\frac{(1-z\,\tilde{z}(t_+))^{1/2}\,(1-z\, \tilde{z}(t_-))^{1/2}}{(1+
\tilde{z}(t_+))^{1/2}\, (1+ \tilde{z}(t_-))^{1/2}}.
\eea

\bfig[htb]
  \begin{center}
   \includegraphics[angle = 0, width = 3.5in, clip = true]{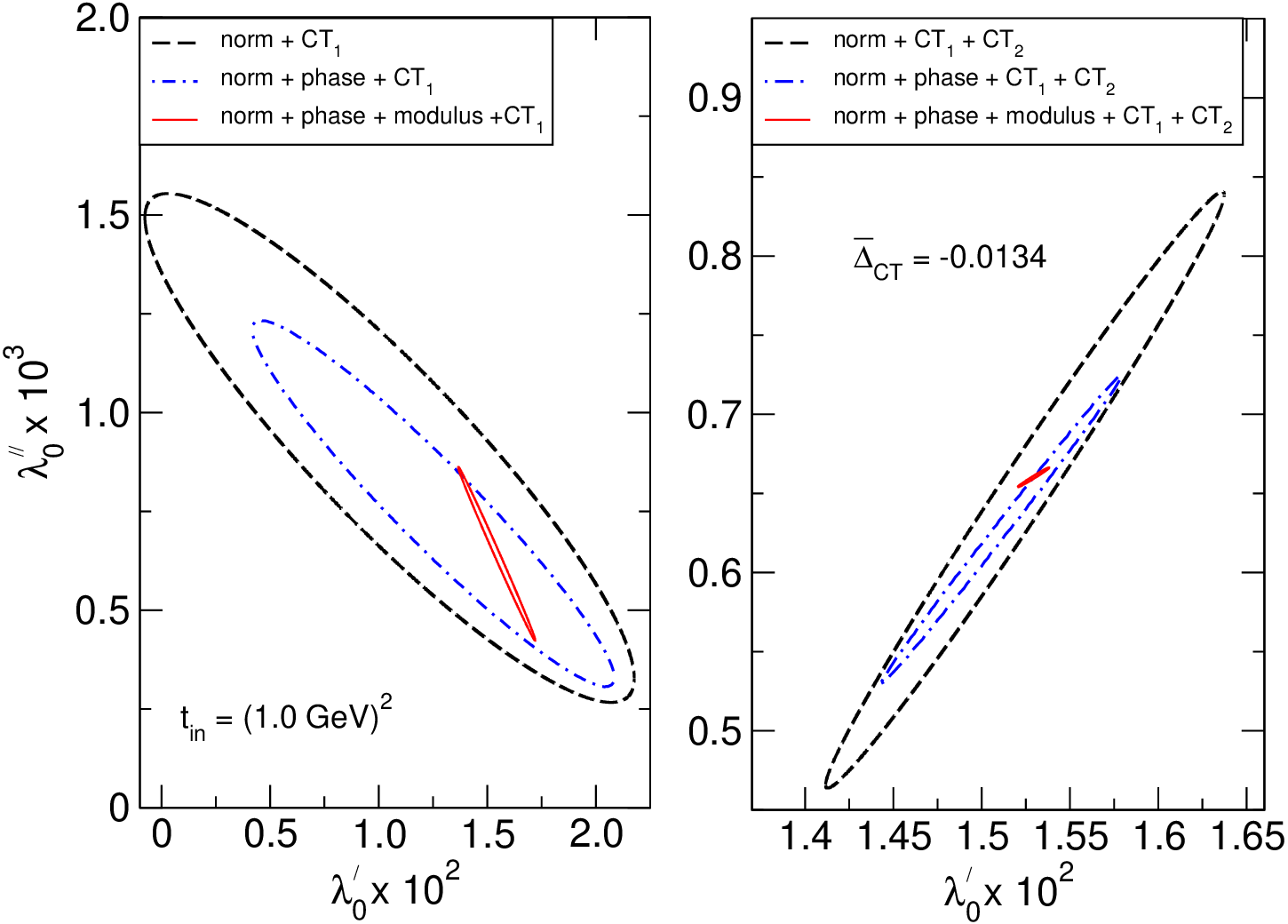}
  \end{center}
  \caption{Allowed regions in the slope-curvature plane of the scalar $K\pi$
form factor, obtained with the techniques of sections \ref{sec:meiman}-\ref{sec:modulus}. 
For explanations see the text.}
\label{fig:fig1}
\efig

In ref. \cite{arXiv:0912.2831} we derived stringent bounds on the slope $\lambda_0'$ and 
curvature $\lambda_0''$, using as input the values of  $f_0(t)$  at $t=0$ and $\ct$, and 
information on the phase and modulus included in the formalism of section \ref{sec:modulus}. 
The additional input from the unitarity cut led to a dramatic improvement of the bounds 
obtained in \cite{arXiv:0903.4297} with the standard unitarity bounds of section \ref{sec:meiman}. 
In \cite{arXiv:0912.2831} we  obtained also a narrow allowed range for the ChPT correction 
$\bar \Delta_{CT}$ at the second Callan-Treiman point $\ctbar$. 

 In the present paper we further illustrate the techniques presented in sections
\ref{sec:meiman}, \ref{sec:phase} and \ref{sec:modulus},  by comparing their
constraining power for various inputs. In fig. \ref{fig:fig1} we present the
allowed domain for the slope and curvature of $f_0(t)$, obtained with the
methods described above: the large ellipse is obtained with standard unitarity
bounds of section \ref{sec:meiman}, the intermediate ellipse includes the phase up
to $\tin=1.0\, {\rm GeV}^2$  with the method of subsection
\ref{sec:lagrange}, and the small ellipse is obtained with the method of section
\ref{sec:modulus}, for the same $\tin$.  The domains shown in the left panel are
 derived using as input the   normalization $f_0(0)=0.962$ and $f_{0}(\Delta_{K
\pi})= 1.193$, those in the right panel use as input also a value at the second
Callan-Treiman point $f_{0}(\bar\Delta_{K\pi}) =1/1.193 + \bar\Delta_{CT}$ for a
certain choice $\bar\Delta_{CT}=-0.0134$.  

As emphasized in section \ref{sec:phase}, the domain obtained with the inclusion 
of the phase should be contained entirely inside the domain obtained with the 
expressions of section \ref{sec:meiman}, since it is related to a minimization on 
a smaller class of admissible functions. This is confirmed by the large and 
intermediate ellipses of fig. \ref{fig:fig1}. On the other hand, the small ellipses 
are given by a minimization on an admissible class, defined in section \ref{sec:modulus},  
which is not {\it a priori} contained in the class defined in section \ref{sec:phase}. 
Therefore, the small ellipses need not to be contained entirely in the intermediate ones. 

In the left panel of fig. \ref{fig:fig1}, the small ellipse is contained inside the other two, 
which means that for points inside this allowed domain all the constraints are fulfilled. 
In the right panel, where we use as input also the value at $\ctbar$, all the domains shrink, 
and the small ellipse has a part situated outside the intermediate one. To satisfy all the 
constraints, one should take the intersection of the small and intermediate ellipses. 

 Clearly, by increasing $\tin$, {\it i.e.} the energy up to which the phase (and
the modulus) are given, the system is more and more constrained and one may
reach a situation when the inputs become inconsistent. This is illustrated in
figs.  \ref{fig:fig2}  and  \ref{fig:fig3}, where we show the configuration of
the allowed domains for two larger values of $\tin$. The large ellipses are the
same in all figures, since they are are independent of $\tin$. As follows from
the arguments of section \ref{sec:phase}, the intermediate ellipses become
gradually smaller when $\tin$ is increased.  In the left panels, obtained with
the normalization at $t=0$ and the value at $\ct$, the small ellipses are
contained inside the intermediate ones, indicating that we can find an allowed
domain that satisfies all the constraints. However,  if we impose also the
constraint at $\ctbar$,  the system becomes over-constrained, and it is
impossible to find a domain that satisfies all the constraints. Indeed, in the
right panels  of figs. \ref{fig:fig2}  and \ref{fig:fig3}, the small ellipses
are not inside the intermediate ones. 

\bfig[htb]
  \begin{center}
   \includegraphics[angle = 0, width =3.5in, clip = true]{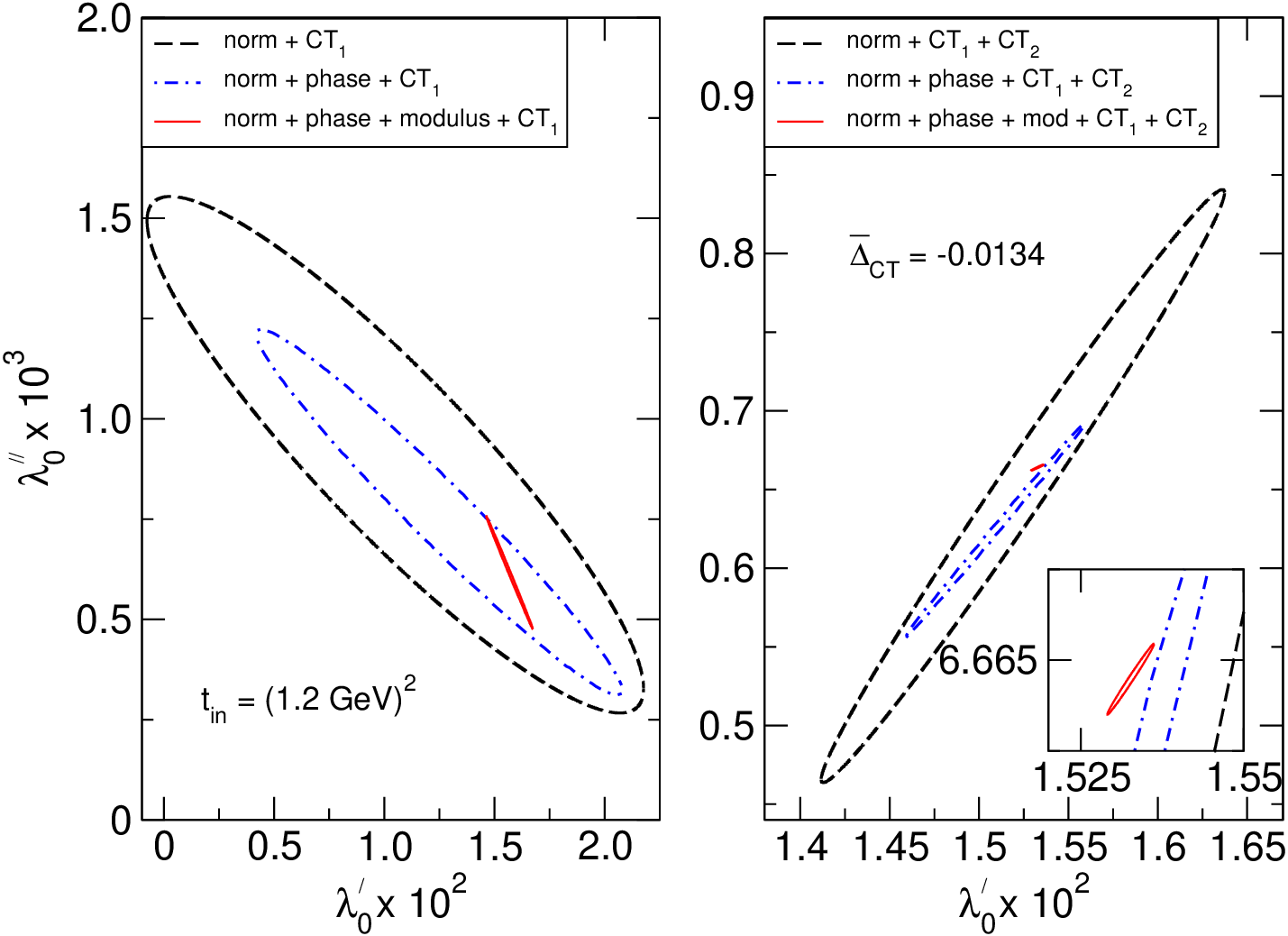}
  \end{center}
  \caption{As in fig. \ref{fig:fig1} for $\tin=(1.2\, {\rm GeV})^2$. The inset zooms in on the smallest ellipse obtained using
the techniques of section \ref{sec:modulus}.}
\label{fig:fig2}
\efig
\bfig[htb]
  \begin{center}
   \includegraphics[angle = 0, width = 3.5in, clip = true]{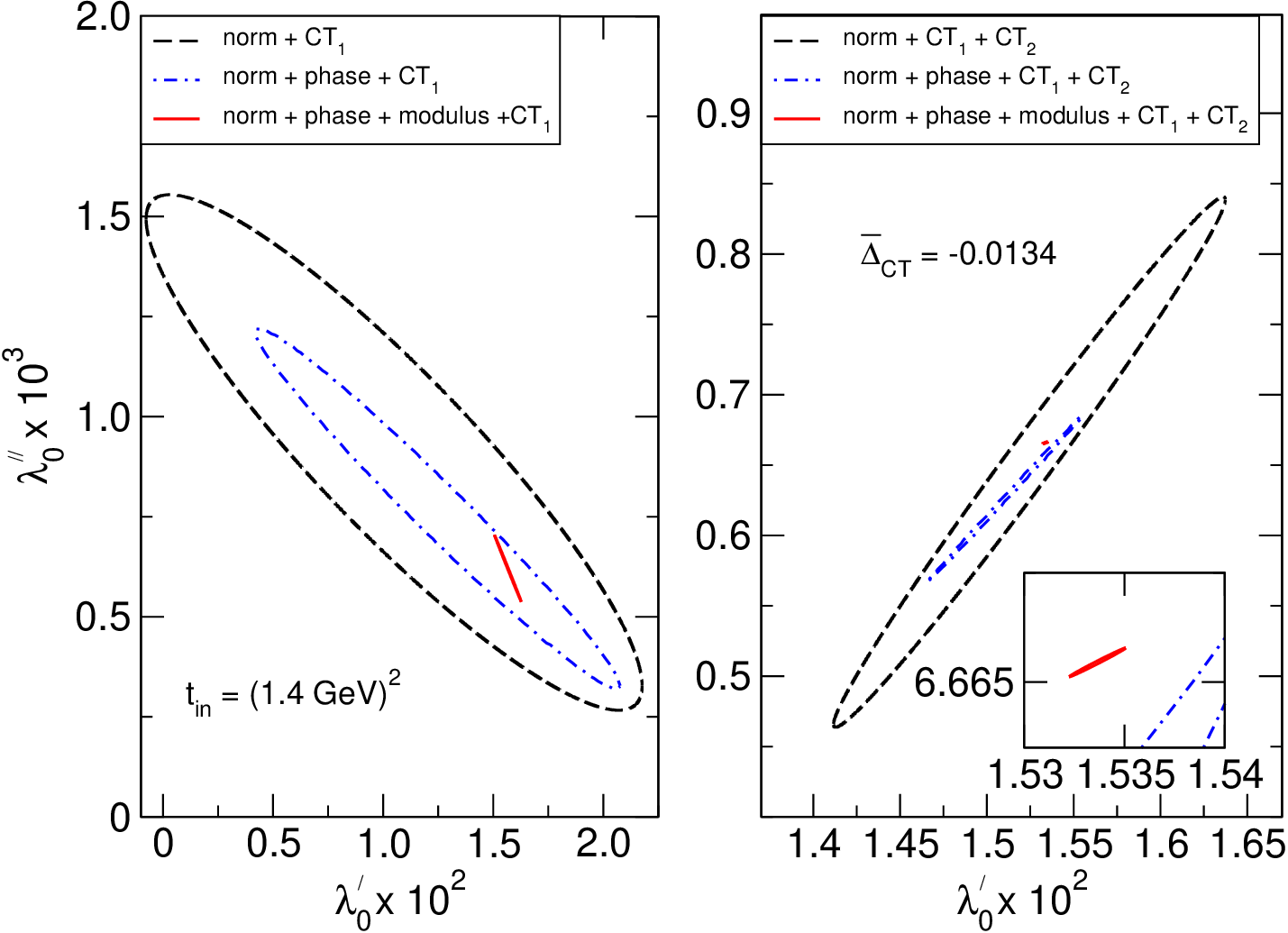}
  \end{center}
  \caption{As in fig. \ref{fig:fig1} for $\tin=(1.4\, {\rm GeV})^2$. The inset zooms in on the smallest ellipse obtained using
the techniques of section \ref{sec:modulus}.}
\label{fig:fig3}
\efig

The same behavior is illustrated  in figs. \ref{fig:comp1} and 
\ref{fig:comp2}, where we show the allowed regions in the slope-curvature plane
for various values of the inelastic threshold $\tin$. The left panels 
show that by increasing $\tin$ the ellipses obtained with the phase
included as in section \ref{sec:phase} become gradually smaller and are contained
one within the other. This is not the case with the ellipses derived with the
phase and modulus included as in section \ref{sec:modulus}. In the particular case
when both $\ct$  and $\ctbar$ are used as input, the ellipses obtained with
various $\tin$ have a zero intersection (see right panel of fig.
\ref{fig:comp2}). This signals an inconsistency of the various quantities used
as input (normalization at $t=0$, values at $\ct$ and $\ctbar$, the phase up to
$\tin$ and the modulus below $\tin$). In particular, the assumption of
neglecting the inelastic effects up to $\tin=(1.4\, {\rm GeV})^2$  may be too
strong.  We note however that the above figures were obtained with the central
values of the inputs described in \cite{arXiv:0912.2831}.  Of course, in
phenomenological analyses one should account for the errors of the various
pieces of the input. We did not consider this aspect here, because our purpose
was only to illustrate the mathematical techniques in a definite
framework. Thus we have provided a concrete illustration of the various
techniques discussed in detail in the previous sections. The phenomenological
implications of the results will be analyzed in a future work.

\bfig[htb]
  \begin{center}
   \includegraphics[angle = 0, width = 3.5in, clip = true]{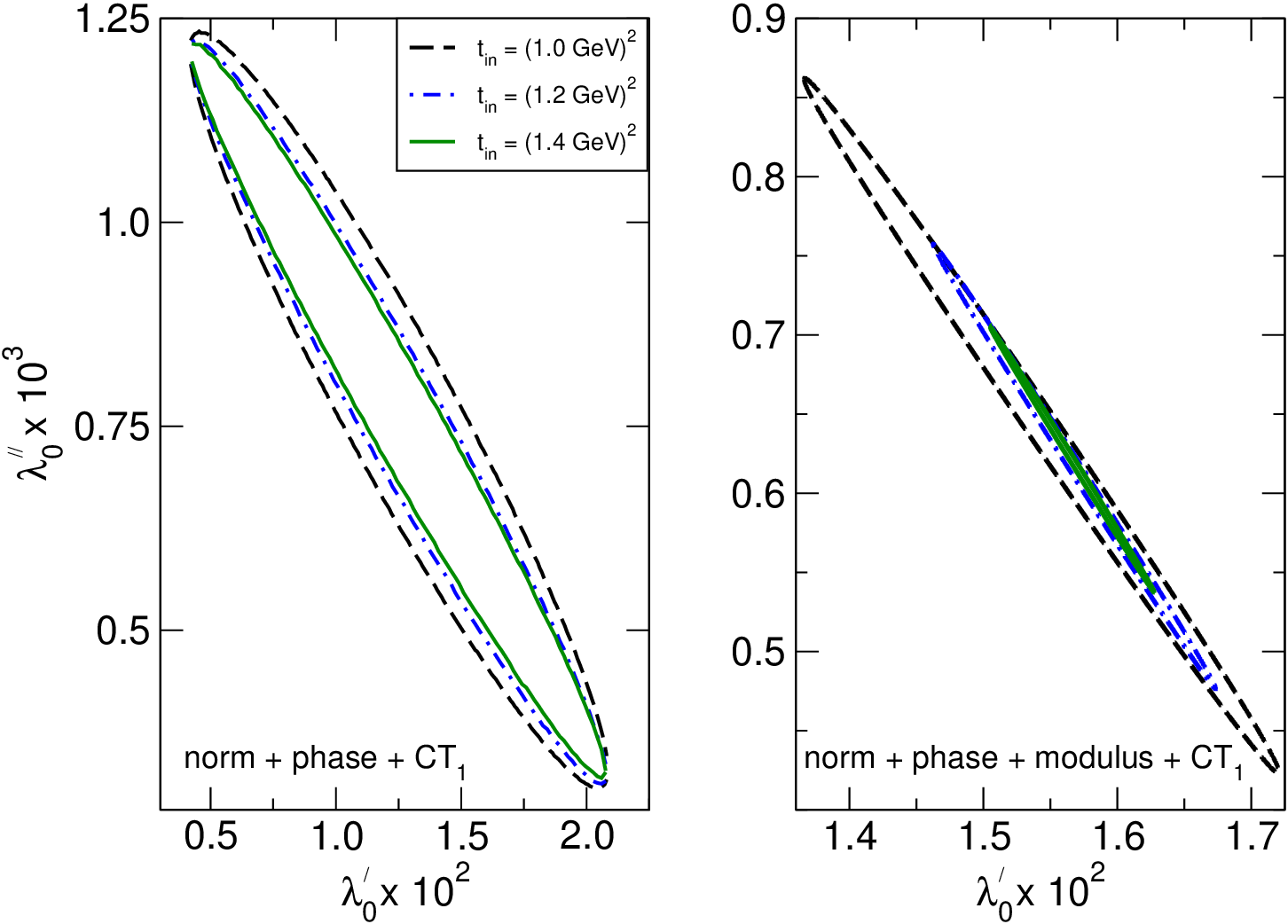}
  \end{center}
  \caption{Allowed region in the slope-curvature plane for various $\tin$ using as 
input the normalization at $t=0$ and the value at $\ct$. Left: phase included as in 
section \ref{sec:phase}. Right: phase and modulus included as in section \ref{sec:modulus}.}
\label{fig:comp1}
\efig

\bfig[htb]
  \begin{center}
   \includegraphics[angle = 0, width = 3.5in, clip = true]{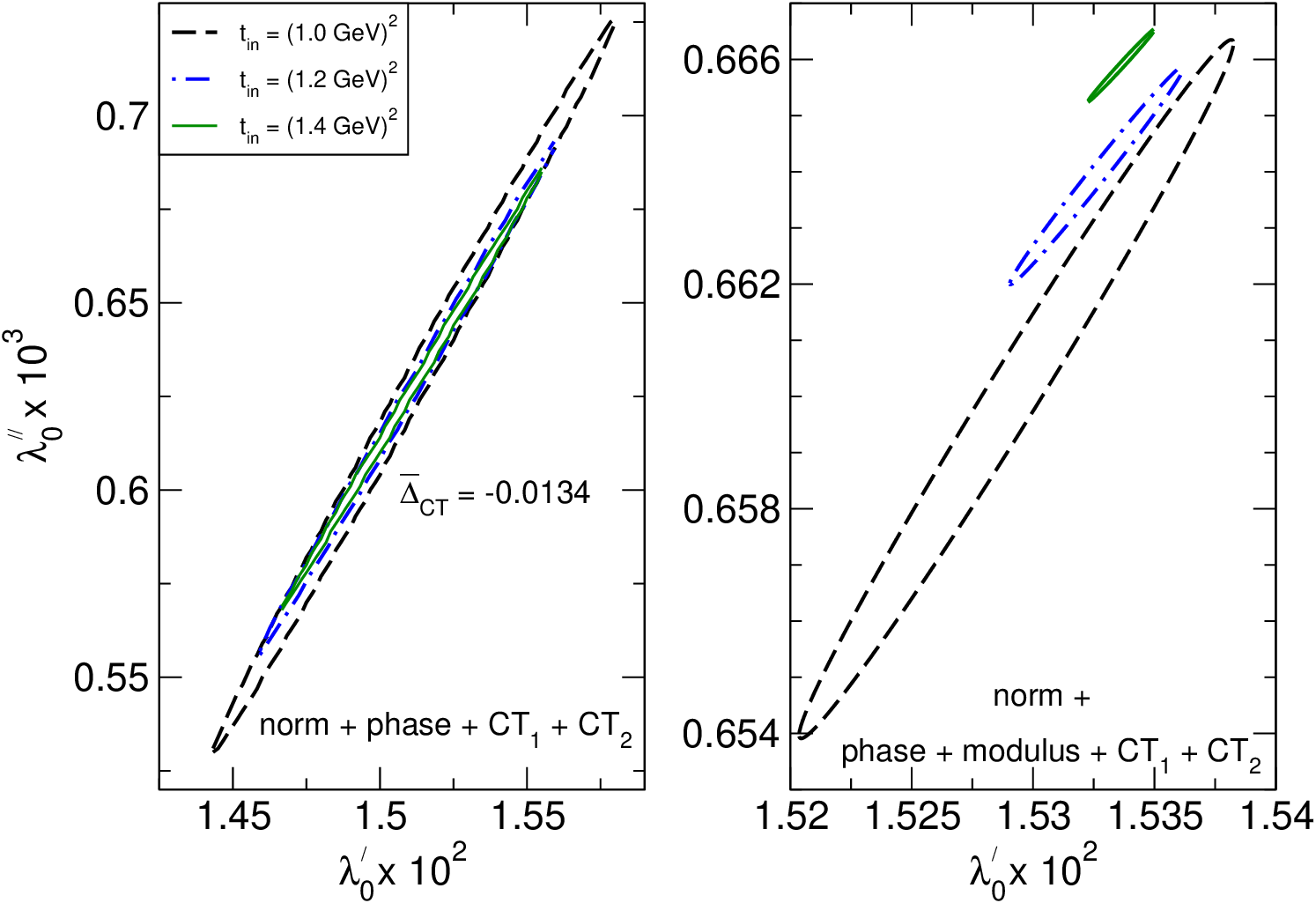}
  \end{center}
  \caption{Allowed region in the slope-curvature plane for various $\tin$ using as input 
the normalization at $t=0$ and the values at $\ct$  and $\ctbar$. Left: phase included 
as in section \ref{sec:phase}. Right: phase and modulus included as in section \ref{sec:modulus}. }
\label{fig:comp2}
\efig

\section{Conclusions}

In this paper we reviewed the method of unitarity bounds, extended in order to 
include information on the phase and modulus of the form factor on the unitarity cut. 
We provided explicit formulas, easily implementable in {\it Mathematica} or $C$ programs, 
for an arbitrary number of derivatives at $t=0$ and an arbitrary number of interior points. 

The method is very suitable for correlating through analyticity various pieces of 
information about the form factors: perturbative QCD, lattice calculations and 
effective theories like ChPT or SCET. It does not depend on specific assumptions 
usually adopted in the standard dispersion relations, like the absence  of zeros, 
or the behavior of the form factor above the inelastic  threshold. 

As shown in section \ref{sec:example}, the  techniques presented here provide strong 
constraints on the shape parameters of the form factors, of interest for the  
parameterizations of the experimental data. The method can be used also to test the 
low energy theorems and put constraints on the higher order corrections of 
ChPT \cite{arXiv:0912.2831}. Moreover, it can be adapted in order to control theoretically 
the truncation error of the experimental parametrization  \cite{CaLeNe,BoCaLe}.

 Another possible application is the detection of the zeros of the form factors.
The problem is of interest  for the Omn\`es representations used recently for
the parametrization of various form factors, which assume that the zeros are
absent. For such a study, one has to assume that $F(t_0)=0$ for a certain
unknown $t_0$, include this condition among the interior constraints
(\ref{eq:cond}) and test the consistency of the inputs, in one of the versions
presented in sections \ref{sec:meiman}-\ref{sec:modulus}. The method then can give
in an unambiguous way the points $t_0$ where zeros are excluded. 

The method of unitarity bounds  proved to be very useful for the description of the 
$B\to D$, $B\to\pi$ or $K\to \pi$ form factors. It can  be applied also to other 
form factors, such as those describing $D\to \pi$ semileptonic decays, or the scalar 
form factors of the pion or kaon,  for which bounds of the type (\ref{eq:I}) can be 
obtained from the dispersion relation of a suitable correlator calculated in perturbative QCD. 
The method is a valuable tool for increasing the precision of the predictions in low energy flavor physics,
which has been discussed here in great detail and generality in an
accessible manner. 

\vskip0.5cm

{\small {\it Acknowledgements.} 
BA thanks DST, Government of India, and the Homi Bhabha Fellowships Council for support. IC acknowledges support from  CNCSIS in the Program Idei, Contract No.
464/2009, and from ANCS, project PN 09370102.}

\end{document}